\documentclass[12pt]{article}
\usepackage[margin=1in]{geometry}
\usepackage{amsmath,amssymb,amsthm}
\usepackage{graphicx}
\usepackage{booktabs}
\usepackage{natbib}
\usepackage{hyperref}
\usepackage{setspace}
\usepackage{tikz}
\usetikzlibrary{positioning, fit, shapes, arrows.meta, shapes.geometric, calc}

\usepackage{mathtools}
\mathtoolsset{showonlyrefs}

\title{A Neural Estimation Framework for Aggregated Relational Data under Intractable Likelihoods}

\author{
  Rowland G. Seymour \\
  School of Mathematics, University of Birmingham
  \and
  Joseph Marsh \\
  School of Mathematics, University of Birmingham
}

\begin{document}

\maketitle

\begin{abstract}
Aggregated relational data (ARD) consists of survey responses to questions of the form ``how many people do you know who~$X$?'' and is widely used in survey statistics for indirect inference about populations and social networks. The dominant ARD inference target is hidden-population size estimation via the Network Scale-Up Method (NSUM), but ARD is also used for personal-network-size estimation, mixing-pattern recovery, and inference about latent network structure. Bayesian inference for ARD almost universally assumes that, conditional on a respondent's degree, the counts reported for different subpopulations are independent. There are, however, reasons to question this assumption, as homophily, latent-space clustering, and imperfect recall may all induce cross-population dependence. We develop a simulation-based neural estimation framework for ARD which requires only a simulator, so it can be applied to generative models whose likelihood cannot be written down or efficiently evaluated. The framework trains a permutation-invariant neural Bayes estimator that returns, for each marginal parameter, a posterior median and a $95\%$ credible interval, by minimising a multi-quantile pinball loss with a cumulative-gap construction that rules out quantile crossing by design. We demonstrate the framework on three structurally distinct intractable extensions of NSUM-style ARD inference: a stochastic block model, a latent-space model, and a recall-subset model. We apply the framework to ARD Household Survey collected in Rwanda. The framework provides inference on any new survey drawn from the training distribution, and extends the reach of ARD modelling to network-structure and cognitive-process assumptions beyond those currently accessible to likelihood-based inference.
\end{abstract}

\medskip
\noindent\textbf{Keywords:} Aggregated relational data; Network scale-up method; hidden population estimation; neural Bayes estimators; simulation-based inference.

\newpage

\section{Introduction}
\label{sec:intro}

Aggregated relational data (ARD) consists of survey responses to questions of the form ``how many people do you know who $X$?'', where $X$ specifies a population of interest such as members of a hidden subgroup, holders of a particular occupation, or people with a specific name \citep{killworth1998estimation, bernard2010counting}. Because the responses summarise the respondent's personal network rather than enumerate it directly, ARD is a workhorse instrument for indirect inference about populations and social networks in settings where direct enumeration is infeasible \citep{mccormick2010many}. The dominant inferential use of ARD is hidden-population size estimation via the Network Scale-Up Method (NSUM) \citep{Lubbers2026}, with applications ranging from people who inject drugs and sex workers to undocumented migrants and victims of human trafficking \citep[see, e.g.,][]{unaids2010guidelines, maltiel2015estimating, NyarkoAgyei2025}. Accurate size estimates in such settings are essential for resource allocation, policy development, and programme evaluation. ARD is also used to estimate personal-network sizes \citep{mccormick2010many}, latent network structure and mixing patterns \citep{zheng2006many}, and barrier and transmission effects in respondents' awareness of their alters \citep{maltiel2015estimating}. The framework we develop applies in principle to all of these inferential uses; our case studies and application focus on the NSUM hidden-population case as the most policy-relevant.

Bayesian inference for ARD is attractive because it propagates uncertainty from individual-level degree parameters through to population-level estimates, and it is the framework within which most modern ARD methods, including modern NSUM extensions, have been formulated \citep{maltiel2015estimating, laga2023correlated}. In practice, essentially all implementations rely on a common simplifying assumption that, conditional on a respondent's personal network size $d_i$, the counts they report for different subpopulations are independent. This assumption is mathematically convenient, in that it yields a tractable Binomial or Poisson likelihood in which each respondent contributes a product of per-population factors, but it is known to be wrong. Several mechanisms induce cross-population dependence in real ARD. Social networks exhibit homophily and block-like community structure \citep{mcpherson2001birds, girvan2002community}, so respondents embedded in one community tend to know disproportionately many people from several populations concentrated there. Acquaintanceship is spatially localised \citep{hoff2002latent}, so populations whose centres are close in some latent social space generate positive correlations among nearby respondents. Respondents do not enumerate all their alters before answering either; they retrieve a random subset and count within it \citep{brewer2000forgetting}, so a forgetful respondent under-reports every population simultaneously.

The standard workaround is to assume the dependence away. The resulting inference is internally well-calibrated but miscalibrated against the true data-generating process. Writing down a correct likelihood function for even a modest generative extension produces an integral with no closed form. A stochastic block model requires marginalising over discrete community labels and a latent graph; a latent-space model requires marginalising over $O(n)$ continuous coordinates with rotation and reflection symmetries; a recall-subset model requires marginalising over $n$ weakly identified per-respondent recall factors. In practice, all three sit beyond what likelihood-based inference can deliver.

In this paper we develop a neural estimation framework for ARD that sidesteps the likelihood entirely. The framework is based on recent developments in simulation-based inference \citep{cranmer2020frontier, sainsbury2024likelihood}. The estimator requires only a simulator, so it admits any generative model for which one can draw pairs of model parameters and simulated data, whatever the analytical status of the likelihood. The framework is the methodological contribution of the paper. The three intractable NSUM extensions that we use to demonstrate it are case studies, chosen to stress the framework in structurally different ways: a discrete latent space with finite-dimensional nuisance (stochastic block model), a continuous high-dimensional latent space with continuous symmetries (latent-space model), and an $n$-dimensional latent space of weakly identified per-respondent parameters (recall-subset model).

Our contributions are as follows. First, we develop an amortised neural estimation framework for Bayesian inference on ARD, combining a permutation-invariant DeepSets encoder \citep{zaheer2017deep} with a multi-quantile head that returns a posterior median and a $95\%$ credible interval for each marginal parameter. The estimator uses a cumulative-gap parameterisation that rules out quantile crossing by construction and is trained by pinball-loss minimisation. Second, we exhibit three intractable extensions of NSUM-style ARD inference and show that, within each trained specification, the same pipeline delivers near-nominal $95\%$ credible-interval coverage on prior-predictive draws. We quantify the cost of the conditional-independence assumption by running MCMC on the standard Binomial-independent NSUM likelihood on data simulated from each of the three generative processes, and we report the resulting miscalibration. Third, we apply the framework to data from the Rwanda Estimating the Size of Populations through a Household Survey.

The remainder of this paper is structured as follows. Section~\ref{sec:background} reviews ARD, NSUM as the dominant ARD application, and the simulation-based tools we use. Section~\ref{sec:framework} develops the neural estimation framework. Section~\ref{sec:models} specifies the three intractable NSUM extensions that we use as case studies. Section~\ref{sec:simulation} reports the simulation study and the MCMC misspecification benchmark. Section~\ref{sec:rwanda} reports the Rwanda application. Section~\ref{sec:discussion} concludes.

\section{Background}
\label{sec:background}

\subsection{Aggregated relational data}
Consider a population of size $N$ that consists of $K' + K$ possibly overlapping subpopulations, where $K'$  and $K$ corresponds to the number of known and hidden populations respectively. The standard ARD structure is a matrix of non-negative integer counts $y_{ik}$ for respondents $i = 1, \ldots, n$, where $y_{ik}$ records respondent $i$'s reported count of contacts in subpopulation $k$. The targets typically partition into known populations whose sizes are verifiable from administrative records (e.g.\ holders of a particular occupation, or people with a specific name) and hidden populations whose sizes are the inferential interest (e.g. people being trafficked, or people who use drugs).

\subsection{The Network Scale Up Method}
NSUM rests on the premise that if personal networks are random samples of the population, the proportion of each respondent's network belonging to a subpopulation estimates that subpopulation's prevalence. The standard generative model takes the form
\begin{equation}
y_{ik} \mid d_i, p_k \sim \mathrm{Binomial}(d_i, p_k),
\label{eq:nsum_binomial}
\end{equation}
where $y_{ik}$ is respondent $i$'s reported count of contacts in subpopulation $k$, $d_i$ is respondent $i$'s personal network size, and $p_k = N_k / N$ is the prevalence of subpopulation $k$ \citep{killworth1998estimation}. Maximising the resulting likelihood with respect to $N_k$, with degrees treated as fixed, yields the basic (or Killworth) estimator
\begin{equation}
\hat{N}_k = \frac{\sum_{i=1}^{n} y_{ik}}{\sum_{i=1}^{n} \hat{d}_i} \times N,
\label{eq:killworth}
\end{equation}
where $\hat{d}_i$ is an estimate of respondent $i$'s personal network size and $N$ is the total population size. Because individual degrees $d_i$ are unobserved, the standard approach uses known populations, that is, subpopulations with established sizes from administrative records, to estimate these parameters.

\subsection{Bayesian NSUM and its structural assumption}
\citet{maltiel2015estimating} developed a Bayesian framework that builds on \eqref{eq:nsum_binomial} and allows for transmission bias (differential awareness of alters' group membership) and barrier effects (non-random mixing between groups). In its simplest form, degrees are assigned a log-normal prior,
\begin{equation}
\log d_i \sim \mathcal{N}(\mu_d, \sigma_d^2),
\label{eq:nsum_degree_prior}
\end{equation}
with $(\mu_d, \sigma_d)$ calibrated empirically to around $\mu_d \approx 5.5$ and $\sigma_d \approx 0.7$ in US settings \citep{zheng2006many, mccormick2010many}. Across \citet{maltiel2015estimating}, \citet{laga2023correlated} and related work, the likelihood factorises as
\begin{equation}
p(Y \mid p_1, \ldots, p_K) = \prod_{i=1}^{n} \prod_{k=1}^{K} p(y_{ik} \mid d_i, p_k). 
\label{eq:factorisation}
\end{equation}
Conditional on $d_i$, the counts $y_{i1}, \ldots, y_{iK}$ are mutually independent. The entire apparatus of MCMC sampling for NSUM exploits this factorisation.

This factorisation is the assumption that each of the three mechanisms described in Section~\ref{sec:intro} violates. When the true data-generating process exhibits cross-population dependence, inference run under \eqref{eq:factorisation} produces a posterior that is internally consistent but miscalibrated against the data-generating process. The inferential target is the population-level vector $\theta = (\mu_d, \sigma_d, p_1, \ldots, p_{K})$, i.e.\ only the hidden-population prevalences

\subsection{Simulation-based neural inference}

Simulation-based inference (SBI) bypasses likelihood evaluation by learning a mapping from data to summaries of the posterior, directly from simulated examples \citep{cranmer2020frontier}. A neural Bayes estimator (NBE) trains a neural network $\hat\theta_\varphi$ to minimise an expected loss over the  joint distribution, averaged over all possible realisations,
\begin{equation}
\hat\varphi = \arg\min_\varphi \, \mathbb{E}_{\theta \sim \pi(\theta)} \, \mathbb{E}_{Y \sim p(Y \mid \theta)} \bigl[ L(\theta, \hat\theta_\varphi(Y)) \bigr],
\label{eq:nbe}
\end{equation}
where the choice of loss $L$ determines the posterior summary that the trained network targets \citep{sainsbury2024likelihood}. Under $L_1$ loss, the population minimiser is the componentwise posterior median; under squared loss, the posterior mean. To quantify uncertainty, we use a generalisation of the standard NBE in which the loss in \eqref{eq:nbe} is the multi-quantile pinball loss \citep{koenker2005quantile}. The pinball loss at level $\tau \in (0, 1)$,
\begin{equation}
\rho_\tau(\theta - \hat q) = \begin{cases} \tau (\theta - \hat q), & \theta \geq \hat q, \\ (1 - \tau)(\hat q - \theta), & \theta < \hat q, \end{cases}
\label{eq:pinball}
\end{equation}
has the well-known property that its population minimiser equals the $\tau$-quantile of the conditional distribution of $\theta$ given $Y$. Training a neural network to minimise the pinball loss summed over a grid of $\tau$ values is therefore Fisher-consistent for the corresponding marginal posterior quantiles, with no tractable likelihood required at any point. We refer to a network so trained as an \emph{interval-NBE}, since at the grid $\mathcal{T} = \{0.025, 0.5, 0.975\}$ it returns a posterior median together with the endpoints of an equal-tailed $95\%$ credible interval.

\section{The Neural Estimation Framework}
\label{sec:framework}

\subsection{Inferential target and conditions of applicability}

The inferential target is the population-level estimand $\theta = (\mu_d, \sigma_d, p_1, \ldots, p_{K})$, i.e.\ only the hidden-population prevalences. Throughout the paper the known-population prevalences are treated as fixed inputs, since they are verifiable from administrative records and serve to anchor each respondent's degree via the Killworth identity~\eqref{eq:killworth}. Each generative model we consider introduces its own nuisance parameter vector $\eta$, whose components vary across surveys and are not of direct interest. The framework handles nuisances by marginalisation during training. The simulator draws $\eta$ afresh from a prior $\pi(\eta)$ for each simulated dataset, and the trained estimator therefore learns a marginal map $Y \mapsto \theta$ that integrates $\eta$ out.

Broadly speaking, the framework applies whenever three conditions hold. First, a simulator is available: one can draw $(\theta, \eta)$ and then $Y \sim p(Y \mid \theta, \eta)$. Second, the estimand $\theta$ is identifiable from the marginal likelihood $p(Y \mid \theta) = \int p(Y \mid \theta, \eta) \pi(\eta) \,\mathrm{d}\eta$. Third, the training prior on $(\theta, \eta)$ covers a large number of prevalence regimes, including the regime from which the real survey is drawn. The framework is the inferential machinery, consisting of a DeepSets encoder feeding an NBE trained on simulated data. It is not related to the modelling specification, which is a choice of generative model together with priors on $\theta$ and $\eta$. The three case studies in Section~\ref{sec:models} are three specifications, and the coverage results in Section~\ref{sec:simulation} demonstrate that the framework is adequate for these three specifications. %They do not, and cannot, demonstrate that the framework is adequate for arbitrary intractable generative models. Each new application requires its own modelling justification, its own choice of nuisance priors. 
The framework's value lies in both making estimator-based inference possible for different kinds of NSUM models, and in providing a pipeline in which new generative assumptions about the underlying data-generating process can be tested, validated, and deployed, where likelihood-based inference is infeasible.

\begin{figure}[htbp]
\centering
% This forces the diagram to fit the width of your text precisely
\resizebox{\textwidth}{!}{%
\begin{tikzpicture}[
    >=Stealth,
    % Reduced global node distance
    node distance=0.8cm and 0.8cm,
    % Styles (slightly adjusted for compactness)
    layerbox/.style={draw, rectangle, minimum height=3cm, minimum width=2cm, align=center, rounded corners, fill=blue!5},
    smallbox/.style={draw, rectangle, minimum height=0.8cm, minimum width=1.2cm, align=center, fill=green!5},
    pool/.style={draw, circle, minimum size=1cm, align=center, fill=orange!10},
    summarybox/.style={draw, rectangle, minimum height=1.5cm, minimum width=1.8cm, align=center, rounded corners, fill=purple!5},
    thick_arrow/.style={->, thick}
]

% 1. Inputs (Data Vectors)
\node[smallbox] (y1) {$y_1$};
\node[smallbox, below=0.4cm of y1] (y2) {$y_2$};
\node[below=0.05cm of y2] (dots) {$\vdots$};
\node[smallbox, below=0.05cm of dots] (yn) {$y_n$};

% 2. DeepSets inner network psi (per-respondent)
\node[smallbox, right=0.7cm of y1] (psi1) {$\psi(y_1)$};
\node[smallbox, right=0.7cm of y2] (psi2) {$\psi(y_2)$};
\node[smallbox, right=0.7cm of yn] (psin) {$\psi(y_n)$};

% Arrows from y_i to psi(y_i)
\draw[thick_arrow] (y1) -- (psi1);
\draw[thick_arrow] (y2) -- (psi2);
\draw[thick_arrow] (yn) -- (psin);

% 3. Mean pool  (matches (1/n) sum in the equation)
\node[pool, right=0.8cm of psi2, yshift=-0.4cm] (sum) {$\tfrac{1}{n}\sum$};

% Arrows from psi(y_i) to mean pool
\draw[thick_arrow] (psi1.east) -- (sum.north west);
\draw[thick_arrow] (psi2.east) -- (sum.west);
\draw[thick_arrow] (psin.east) -- (sum.south west);

% 4. Encoder output
\node[summarybox, right=0.8cm of sum] (summary) {Summary\\Vector\\(Size: 64)};
\draw[thick_arrow] (sum) -- (summary);

% Draw bounding box around DeepSets Encoder
\node[draw, dashed, thick, rounded corners, fit=(y1) (yn) (psi1) (psin) (sum) (summary), inner sep=10pt, label={[font=\bfseries]above:DeepSets Encoder}] (deepsets) {};

% 5. Hidden Layer 1
\node[layerbox, right=1.2cm of summary] (h1) {Hidden Layer 1\\(Width: 128)\\+ ReLU};
\draw[thick_arrow] (summary) -- (h1);

% 6. Hidden Layer 2
\node[layerbox, right=0.7cm of h1] (h2) {Hidden Layer 2\\(Width: 128)\\+ ReLU};
\draw[thick_arrow] (h1) -- (h2);

% 7. Output Layer
\node[layerbox, right=1cm of h2, fill=red!5, minimum height=3.5cm, minimum width=2.5cm] (out) {Output Layer\\(Size: $3d$)\\[8pt] \footnotesize For each parameter:\\[2pt] $\hat{\theta}_{0.5}$ (Median)\\$\hat{\theta}_{0.025}$ (2.5\%)\\$\hat{\theta}_{0.975}$ (97.5\%)};
\draw[thick_arrow] (h2) -- (out);

% 8. Outer-network label spanning h1, h2, out
\node[above=0.35cm of h2, font=\bfseries] {Outer Network $\phi$};

\end{tikzpicture}%
} % End of resizebox
\caption{Architecture of the Neural Bayes Estimator.}
\label{fig:nbe_architecture}
\end{figure}
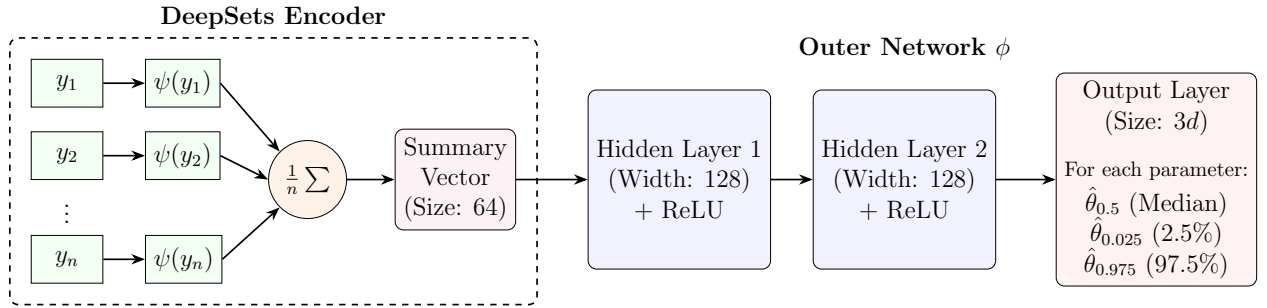

\subsection{Permutation-invariant architecture}

ARD has a natural exchangeability structure, where the order of respondents carries no information. Standard neural architectures do not respect this invariance and can in principle learn spurious patterns from respondent ordering. We employ a DeepSets architecture \citep{zaheer2017deep}, which guarantees permutation invariance through the structure
\begin{equation}
f(\{y_1, \ldots, y_n\}) = \phi\!\left( \frac{1}{n} \sum_{i=1}^{n} \psi(y_i) \right),
\label{eq:deepsets}
\end{equation}
where $\psi$ is an inner network that processes each respondent's data independently, the mean pool aggregates over respondents, and $\phi$ is an outer network that maps the aggregated representation to the quantity of interest. The input features for respondent $i$ are the log-transformed counts $\log(y_{ik}+1)$ for all $K$ populations, concatenated with the known-population prevalences. Mean pooling produces a fixed-dimensional summary regardless of sample size $n$, so the same estimator can in principle handle surveys of different sizes. Figure \ref{fig:nbe_architecture} shows a diagram of the architecture. 

We use the DeepSets encoder rather than more expressive alternatives such as set transformers because it is the  simplest architecture that respects the required invariance, and the ARD summary statistics of interest are essentially moment-like, and mean-pooled DeepSets are well-suited to capturing such summaries. 

\subsection{Multi-quantile head with cumulative-gap construction}
\label{sec:cumgap}

The outer network $\phi$ in \eqref{eq:deepsets} produces, for each marginal parameter, three quantile estimates at $\mathcal{T} = \{0.025, 0.5, 0.975\}$. Each quantile level is served by its own head, which consists of a separate two-hidden-layer multi layer perceptron $v_{\tau_t}: \mathbb{R}^{d_e} \to \mathbb{R}^d$, where $d_e$ is the encoder dimension and $d$ is the dimension of $\theta$. The heads are not conditioned on $\tau$. Instead, the three estimates are stitched together through a cumulative-gap construction. We let $\tau_1 < \tau_2 < \tau_3$ denote the three quantile levels, and write $g(\cdot) = \mathrm{softplus}(\cdot)$. The estimator then returns
\begin{equation}
\hat q_{\tau_1}(Y) = v_{\tau_1}(\Psi(Y)), \qquad \hat q_{\tau_t}(Y) = v_{\tau_1}(\Psi(Y)) + \sum_{j=2}^{t} g(v_{\tau_j}(\Psi(Y))), \quad t = 2, 3,
\label{eq:cumgap}
\end{equation}
where $\Psi(Y) = n^{-1} \sum_{i=1}^{n} \psi(y_i)$ is the pooled DeepSets encoding. Because $g$ is positive, the construction guarantees $\hat q_{\tau_1}(Y) \leq \hat q_{\tau_2}(Y) \leq \hat q_{\tau_3}(Y)$ for every input, every parameter coordinate, and every choice of network weights, so quantile crossing is impossible.

A more flexible alternative is a $\tau$-conditional network $\hat q_\tau(Y) = \phi(\Psi(Y), \tau)$ trained over $\tau \sim \mathrm{Uniform}(0, 1)$, with monotonicity in $\tau$ enforced by constraining the weights on the $\tau$ pathway to be positive \citep{cannon2018non}. We tried this design and observed collapse on the NSUM simulators in Section~\ref{sec:models}. For most test datasets the network produced $\hat q_{0.025}(Y) \approx \hat q_{0.5}(Y) \approx \hat q_{0.975}(Y)$, yielding intervals of essentially zero width. The cumulative-gap construction \eqref{eq:cumgap} replaces the soft monotonicity constraint with a hard one and the failure mode disappears: each gap term $g(v_{\tau_j}(Y))$ is bounded below by $\log 2$ at initialisation, so the network must work to make intervals tight rather than to make them wide.

\subsection{Training protocol}

Training minimises the multi-quantile pinball loss summed across the three quantile levels in $\mathcal{T}$ and the $d$ marginal parameters,
\begin{equation}
\hat\varphi = \arg\min_\varphi \, \mathbb{E}_{\theta \sim \pi(\theta)} \, \mathbb{E}_{Y \sim p(Y \mid \theta)} \sum_{t=1}^{3} \sum_{j=1}^{d} \rho_{\tau_t}\bigl(\theta_j - \hat q_{\tau_t}(Y; \varphi)_j\bigr),
\label{eq:multi_pinball_loss}
\end{equation}
where $\rho_\tau$ is the pinball loss in \eqref{eq:pinball}. The expectation is approximated by Monte Carlo over $K = 5{,}000$ prior-predictive draws $(\theta_k, Y_k)$, refreshed during training. Throughout the paper we use an inner network with two hidden layers of 128 units and ReLU activations, an encoding dimension of 64, and an outer network (per quantile head) with two hidden layers of 128 units. Training uses the Adam optimiser with default learning rate over 200 epochs, with a batch size of 32 and early stopping if validation risk fails to improve over 5 consecutive epochs. Total wall-clock training time on a single CPU core is approximately 25 minutes per simulator. All code is implemented in \textsf{Julia} using \textsf{NeuralEstimators.jl} and \textsf{Flux.jl}.

Nuisance priors are specified for each generative model in Section~\ref{sec:models}. They are chosen to be weakly informative over the range of regimes a practitioner would plausibly consider, and they are deliberately wider than would be appropriate for any single survey. The estimator is trained once for all surveys drawn from the prior support, so the prior should cover the union of plausible regimes rather than the intersection.

\subsection{Validation via empirical interval coverage}

Because the trained estimator is a learned approximation, its calibration must be verified. The natural diagnostic for an interval estimator is empirical coverage of the credible interval on a held out test set. For each case study we draw $N$ parameter vectors $\theta^{(s)}$ from the prior, simulate datasets $Y^{(s)} \sim p(Y \mid \theta^{(s)})$, evaluate the trained estimator to obtain $(\hat q_{0.025}^{(s)}, \hat q_{0.5}^{(s)}, \hat q_{0.975}^{(s)})$, and record, for each marginal parameter $j$, the empirical coverage
\begin{equation}
\widehat C_{95}(j) = \frac{1}{N} \sum_{s=1}^{N} \mathbb{I}\!\left( \hat q_{0.025}(Y^{(s)})_j \leq \theta^{(s)}_j \leq \hat q_{0.975}(Y^{(s)})_j \right),
\label{eq:empirical_coverage}
\end{equation}
together with the mean interval width, where $\mathbb{I}(\cdot)$ denotes the standard indicator function. Under ideal calibration $\widehat C_{95}(j) \approx 0.95$ where values below indicate intervals that are too narrow (over-confidence) and values above indicate intervals that are too wide (under-confidence). Empirical coverage is a single scalar with an unambiguous interpretation, in contrast to rank-based diagnostics in which over-confidence and bias are entangled in a single visual signature \citep{talts2018validating}.

A useful feature of the interval-NBE is that it decouples two diagnostic questions that can fail independently. The first is whether the median is accurate, summarised by the bias, mean absolute error, and Pearson correlation of $\hat q_{0.5}$ against the truth. The second is whether the uncertainty is well-calibrated, summarised by $\widehat C_{95}$ and the mean interval width. Under a structural under-identification the framework does not invent information the data do not contain; instead, the credible interval correctly widens, as we show in the Supplementary Material for the recall-subset case. Limitations of the framework, including its dependence on the training prior and the need to retrain when the survey design changes substantially, are discussed in Section~\ref{sec:discussion}.

\section{Three Intractable Extensions of NSUM}
\label{sec:models}

We now specify the three case studies. In each case we give the generative model, identify the nuisance parameters over which training marginalises, and state why the marginal likelihood is intractable. All three models share the same population-level estimand $\theta = (\mu_d, \sigma_d, p_1, \ldots, p_{K})$, and all three treat known-population prevalences $p^{\mathrm{known}}$ as fixed inputs rather than parameters to be estimated. Known populations enter each simulator the same way they enter standard NSUM: respondent-level counts to populations of established size pin down each respondent's degree $d_i$ via the Killworth identity~\eqref{eq:killworth}. For the simulation study we use priors
\begin{align}
\mu_d \sim \mathrm{Uniform}(5, 6.5), \quad \sigma_d \sim \mathrm{Uniform}(0.3, 1.0), \quad p_1, \ldots, p_K \overset{\mathrm{iid}}{\sim} \mathrm{Beta}(2, 50),
\label{eq:main_priors}
\end{align}
chosen to cover standard NSUM regimes \citep{mccormick2010many}.

\subsection{Stochastic block model NSUM}
\label{sec:sbm}

The first model we demonstrate on is a stochastic block model. Respondents are embedded in a latent stochastic block model with $B$ communities, in which within-block ties are more probable than between-block ties and hidden populations are non-uniformly distributed across blocks. Both features are designed to break the conditional-independence assumption of \eqref{eq:factorisation}. Each respondent $i$ belongs to a latent block $b_i \in \{1, \ldots, B\}$ with probabilities $\pi = (\pi_1, \ldots, \pi_B)$. Let $\rho \geq 1$ be the within-to-between edge-rate ratio: a tie between two members of the same block is $\rho$ times more likely than a tie between members of different blocks, with $\rho = 1$ recovering the unstructured Binomial NSUM model. Combining this preference with the block sizes $\pi$, the probability that a randomly chosen tie from a block-$b$ respondent terminates in block $b'$ is
\begin{equation}
M_{bb'} = \frac{\pi_{b'} \, w_{bb'}}{\sum_{b''} \pi_{b''} \, w_{bb''}},
\qquad w_{bb'} = \begin{cases} \rho, & b = b', \\ 1, & b \neq b', \end{cases}
\label{eq:sbm_mixing}
\end{equation}
i.e.\ block sizes weighted by the within-vs-between preference and row-normalised so that $\sum_{b'} M_{bb'} = 1$. Hidden populations are distributed across blocks via a matrix $\omega \in [0,1]^{B \times K}$ whose columns sum to one, yielding block-conditional prevalences $\phi_{bk} = \min(p_k \omega_{bk} / \pi_b, 1)$ and effective block-level prevalences $\tilde p_{bk} = \sum_{b'} M_{bb'} \phi_{b'k}$. Counts then follow
\begin{equation}
y_{ik} \mid b_i, d_i \sim \mathrm{Binomial}(d_i, \tilde p_{b_i, k}),
\label{eq:sbm_likelihood}
\end{equation}
with known-population counts drawn from the standard model. The latent block labels $b_i$ are unobserved, and the nuisance vector is $\eta^{\mathrm{SBM}} = (\pi, \rho, \omega)$ with priors $\pi \sim \mathrm{Dir}(2\cdot \mathbf{1}_B)$, $\rho \sim \mathrm{Uniform}(1, 20)$, $\omega_{\cdot k} \sim \mathrm{Dir}(0.5 \mathbf{1}_B)$ for $k = 1, \ldots, K$, and $B = 5$. The nuisance prior includes the degenerate case $\rho = 1$, in which SBM reduces to the Binomial NSUM model. Across the three intractable extensions we use the natural sampling distribution for each generative mechanism: Binomial for the SBM, where the marginalised mixing matrix still yields per-block tie probabilities in $[0, 1]$; and Poisson for the latent-space and recall-subset models, where the rate scaling by $W_{ik}$ and $q_i$ respectively can exceed unity and a Binomial would require an artificial clamp.

The full generative process for respondent $i$ is
\begin{align}
b_i &\sim \mathrm{Categorical}(\pi), \\
\log d_i &\sim \mathcal{N}(\mu_d, \sigma_d^2), \\
y_{ik} \mid b_i, d_i &\sim \mathrm{Binomial}(d_i, \tilde p_{b_i, k}), \quad k = 1, \ldots, K,
\end{align}
where $\tilde p_{bk}$ is the block-$b$ effective prevalence. The joint density of the observed counts and latent block labels factorises across respondents,
\begin{equation}
p(Y, b \mid d, \theta, \eta^{\mathrm{SBM}}) = \prod_{i=1}^{n} \pi_{b_i} \prod_{k=1}^{K} \mathrm{Binomial}(y_{ik}; d_i, \tilde p_{b_i, k}).
\label{eq:sbm_joint}
\end{equation}
The observed-data likelihood marginalises over the latent block labels,
\begin{equation}
p(Y \mid d, \theta, \eta^{\mathrm{SBM}}) = \prod_{i=1}^{n} \sum_{b=1}^{B} \pi_b \prod_{k=1}^{K} \mathrm{Binomial}(y_{ik}; d_i, \tilde p_{bk}).
\label{eq:sbm_marginal}
\end{equation}
The posterior of inferential interest is
\begin{equation}
p(\theta \mid Y) \propto p(\theta) \int p(Y \mid \boldsymbol d, \theta, \eta^{\mathrm{SBM}}) \, p(\boldsymbol d \mid \theta) \, p(\eta^{\mathrm{SBM}}) \, \mathrm{d} \boldsymbol d \, \mathrm{d} \eta^{\mathrm{SBM}},
\label{eq:sbm_posterior}
\end{equation}
where $\boldsymbol d = (d_1, \ldots, d_n)$ collects the respondent degrees and $\eta^{\mathrm{SBM}} = (\pi, \rho, \omega)$ collects the structural nuisance: the $B$-vector of block proportions $\pi$ (on the $(B-1)$-simplex), the within-to-between rate ratio $\rho$, and the $B \times K$ matrix of block-conditional prevalences $\omega$ (each column on the $(B-1)$-simplex). With $B = 5$ and $K = 3$ the nuisance integral is $\big(n + (B-1) + 1 + K(B-1)\big) = (n + 17)$-dimensional, with no closed form. The simulation-based framework sidesteps it entirely: the simulator draws $\eta^{\mathrm{SBM}}$ afresh per dataset, and the estimator learns the marginal map $Y \mapsto \theta$ directly.

\subsection{Latent-space NSUM}
\label{sec:latent_space}

The second model introduces spatial heterogeneity. Respondents and hidden populations live in a shared two-dimensional latent social space, capturing the intuition that acquaintanceship is spatially localised: a respondent near a population's centre knows disproportionately many of its members, and populations whose centres are close induce positive cross-population dependence among nearby respondents. Each respondent has an unobserved position $z_i \in \mathbb{R}^2$ and each hidden population has an unobserved centre $\mu_k \in \mathbb{R}^2$,
\begin{equation}
z_i \overset{\mathrm{iid}}{\sim} \mathcal{N}(\boldsymbol{0}, I_2), \qquad \mu_k \overset{\mathrm{iid}}{\sim} \mathcal{N}(\boldsymbol{0}, \sigma_\mu^2 I_2).
\label{eq:latent_positions}
\end{equation}
Respondent $i$'s per-tie rate for population $k$ is scaled by a Gaussian activation
\begin{equation}
W_{ik} = \frac{\exp(-\|z_i - \mu_k\|^2 / (2 \tau^2))}{\bar{W}_{\cdot k}}, \qquad \bar{W}_{\cdot k} = \frac{1}{n} \sum_{i=1}^{n} \exp(-\|z_i - \mu_k\|^2 / (2 \tau^2)),
\label{eq:latent_kernel}
\end{equation}
where $\tau > 0$ is a global bandwidth, and where the column normalisation ensures that the realised marginal prevalence matches the target $p_k$. Counts to hidden populations follow $y_{ik} \sim \mathrm{Poisson}(d_i p_k W_{ik})$, and counts to known populations follow $y_{ik} \sim \mathrm{Poisson}(d_i p^{\mathrm{known}}_k)$, with known populations assumed diffuse in latent space. The nuisance vector is $\eta^{\mathrm{LS}} = (\tau, \sigma_\mu, \boldsymbol{z}, \boldsymbol{\mu})$, where $\boldsymbol{z} = (z_1, \ldots, z_n)$ collects the respondent latent positions and $\boldsymbol{\mu} = (\mu_1, \ldots, \mu_K)$ collects the population centres, with priors $\tau \sim \mathrm{Uniform}(0.3, 2.0)$ and $\sigma_\mu \sim \mathrm{Uniform}(0.5, 2.0)$.

The marginal likelihood is the $(2n + 2 K)$-dimensional integral
\begin{equation}
p(Y \mid \theta, \tau, \sigma_\mu) = \int p(Y \mid \theta, \boldsymbol{z}, \boldsymbol{\mu}, \tau) \prod_i \mathcal{N}(z_i \mid 0, I_2) \prod_k \mathcal{N}(\mu_k \mid 0, \sigma_\mu^2 I_2) \,\mathrm{d} \boldsymbol{z} \,\mathrm{d} \boldsymbol{\mu},
\label{eq:latent_marginal}
\end{equation}
which has no analytical reduction and whose integrand factorises over neither $i$ nor $k$. 

\subsection{Recall-subset NSUM}
\label{sec:recall}

The recall-subset extension is motivated by cognitive models of how respondents actually answer ``how many $X$ do you know?'' Enumerating all $d_i$ alters is known to exceed working-memory limits for typical personal-network sizes, and respondents instead retrieve a random subset of their alters and count within it \citep{killworth1998estimation}. The fraction retrieved varies by respondent. Each respondent has an unobserved recall fraction $q_i \in (0, 1)$ drawn from a Beta distribution,
\begin{equation}
q_i \overset{\mathrm{iid}}{\sim} \mathrm{Beta}(m_q \kappa_q, (1 - m_q) \kappa_q),
\label{eq:recall_qi}
\end{equation}
where $m_q \in (0, 1)$ is the mean and $\kappa_q > 0$ the concentration parameter. Observed counts are Poisson with a rate thinned by $q_i$,
\begin{equation}
y_{ik} \mid d_i, q_i \sim \mathrm{Poisson}(d_i p_k q_i),
\label{eq:recall_likelihood}
\end{equation}
applied identically to known and hidden populations. The same $q_i$ multiplies every population for respondent $i$, so a forgetful respondent under-reports every population simultaneously. This produces both overdispersion relative to Poisson and positive within-respondent cross-population covariance. The nuisance vector is $\eta^{\mathrm{RS}} = (m_q, \kappa_q, \boldsymbol{q})$, where $\boldsymbol{q} = (q_1, \ldots, q_n)$ collects the respondent recall fractions, with priors $m_q \sim \mathrm{Uniform}(0.3, 0.9)$ and $\kappa_q \sim \mathrm{Uniform}(2, 50)$. These ranges are deliberately broad. The support on $m_q$ spans the empirically observed range of recall fractions reported in the cognitive-survey literature, from heavily forgetful respondents (under one-third recall) to near-complete recall, and the support on $\kappa_q$ covers both highly heterogeneous regimes ($\kappa_q \approx 2$, see the Supplementary Material) and near-uniform recall ($\kappa_q = 50$). A practitioner with auxiliary information about respondent recall behaviour in their target population can tighten these ranges; the framework is otherwise agnostic to the specific choice.

The observed likelihood marginalises over the $n$ per-respondent recall fractions,
\begin{equation}
p(Y \mid \theta, m_q, \kappa_q) = \prod_{i=1}^{n} \int_0^1 \left[ \prod_k \mathrm{Poisson}(y_{ik}; d_i p_k q_i) \right] \mathrm{Beta}(q_i \mid m_q \kappa_q, (1 - m_q) \kappa_q) \,\mathrm{d}q_i,
\label{eq:recall_marginal}
\end{equation}
which is a product of $n$ one-dimensional integrals with no closed form (the Poisson-Beta mixture is not conjugate). Each $q_i$ is weakly identified because a single respondent's handful of counts carries only a noisy signal about their own recall fraction, so the joint posterior over $\boldsymbol{q}$ is diffuse, high-dimensional, and tightly correlated with $\theta$ through the shared $p_k$'s. The framework sidesteps the integral entirely by sampling $(m_q, \kappa_q, \boldsymbol{q})$ afresh per simulated dataset and learning the marginal map.

\section{Simulation Study}
\label{sec:simulation}

We evaluate the framework on data simulated from each of the three models of Section~\ref{sec:models} in turn. All simulations use $n = 500$ respondents, $K_\mathrm{known} = 3$ known populations at prevalences $(0.02, 0.015, 0.01)$, and $K = 3$ hidden populations. The nuisance parameters are drawn fresh per dataset from the priors in Section~\ref{sec:models}. An interval-NBE is trained once per model and evaluated on $N = 1{,}000$ independent prior-predictive draws. Hidden prevalences are reported on the log scale which is the same parameterisation the estimator is trained on.

Section~\ref{sec:hmc_comparison} compares the framework against HMC on the Binomial-independent NSUM across both a well-specified positive control (data simulated from standard NSUM, both estimators well-specified) and the three misspecified DGPs of Section~\ref{sec:models} (HMC forced onto the wrong likelihood). Detailed per-DGP performance of the interval-NBE on each of the three intractable DGPs, together with the cross-population correlation diagnostics that confirm each simulator produces its intended structural feature, and a sensitivity analysis under prior misspecification, are reported in the Supplementary Material.

Training each interval-NBE on $K = 5{,}000$ simulated datasets of $n = 500$ respondents took approximately 25 minutes on a single CPU core. After training, inference takes under a millisecond per survey dataset. The full coverage check on $N = 1{,}000$ test datasets completes in seconds once the estimator is trained. Equivalent MCMC-based inference on the simpler Binomial-independent NSUM model runs at minutes per dataset, and on any of the three intractable models is computationally impractical or not possible.

\subsection{Comparison with HMC on the Binomial-independent NSUM}
\label{sec:hmc_comparison}

To quantify the cost of the conditional-independence assumption, we compare the framework interval-NBE against HMC fits of the Binomial-independent NSUM model~\eqref{eq:nsum_binomial} under two regimes. In the well-specified regime the data themselves are simulated from \eqref{eq:nsum_binomial}, so both the NBE and MCMC based estimators target the same generative process; this is the positive control. In the misspecified regime the data are simulated from one of the three intractable DGPs of Section~\ref{sec:models}, so HMC is fitting the wrong likelihood while the framework interval-NBE is trained on a simulator that matches the DGP.

For each of the four regimes (standard, SBM, latent, and recall) we draw 200 parameter vectors $(\theta^{(s)}, \eta^{(s)})$ from the same priors used to train the corresponding interval-NBE. For each draw we simulate a survey of $n = 500$ respondents and fit the Binomial-independent NSUM model by HMC using four chains of 2{,}000 iterations (1{,}000 warmup). Sampling is performed in \texttt{Turing.jl} using the No-U-Turn Sampler (NUTS), so the leapfrog step size and a diagonal mass matrix are tuned automatically during warmup by dual averaging targeting an average acceptance probability of $0.80$. Coverage of central credible intervals at the 50\,\%, 80\,\%, and 95\,\% nominal levels is recorded. The same 200 datasets per regime are then evaluated by the interval-NBE, and the empirical 95\,\% credible-interval coverage is recorded for like-for-like comparison. Priors on $\theta$ are matched between the two estimators so that any coverage difference is attributable to the likelihood rather than to prior mismatch.

\begin{figure}[t]
\centering
\includegraphics[width=\textwidth]{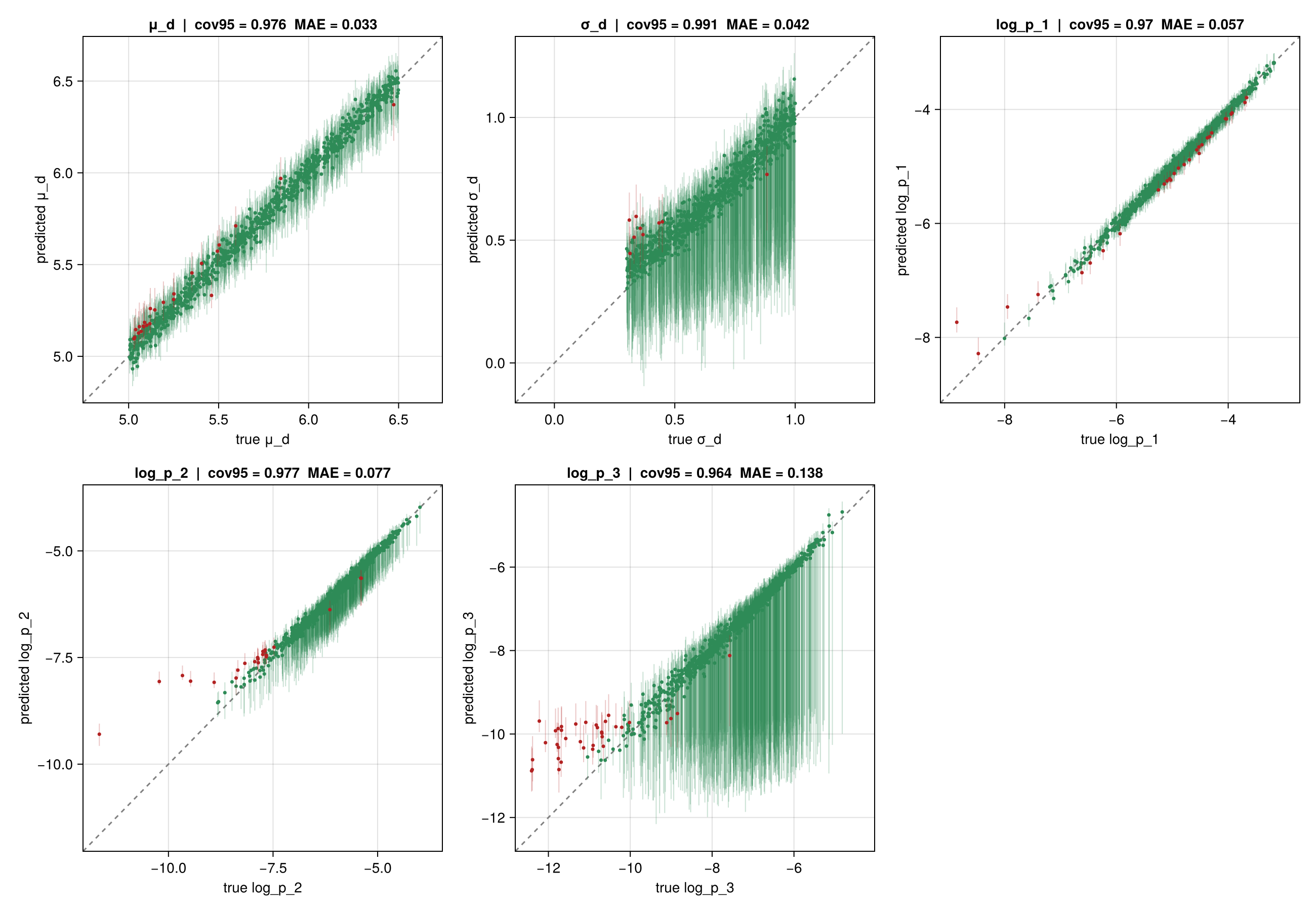}
\caption{Interval-NBE predictions on $N = 1{,}000$ held-out standard-NSUM test datasets. Vertical bars are $95\%$ credible intervals. Points are coloured green when the true value is contained in the interval and red otherwise.}
\label{fig:sim_standard_intnbe}
\end{figure}

Figure~\ref{fig:sim_standard_intnbe} shows the interval-NBE scatter on the well-specified standard-NSUM regime, alongside the per-parameter $95\%$ credible intervals and empirical coverage. Both diagnostics behave as expected, with some discrepancy at the lower end of $p_3$, however as the true value is near zero, there is a difficult parameter to learn.  The median tracks the truth tightly and the $95\%$ interval covers near-nominally for every parameter. Together with Table~\ref{tab:mcmc_nbe_coverage}, this confirms that the framework matches a likelihood-based competitor when both target the same generative process, before we go on to the regimes where likelihood-based inference is no longer well-specified.

Table~\ref{tab:mcmc_nbe_coverage} reports the empirical coverage of basic NSUM (HMC on the Binomial-independent likelihood~\eqref{eq:nsum_binomial}) and the interval-NBE across all four regimes. The first row group, the well-specified standard-NSUM positive control, confirms that basic NSUM and the interval-NBE behave equivalently when the likelihood is correct, with coverage at all three nominal levels close to target and the two estimators agreeing to within Monte Carlo noise. In the three misspecified regimes the picture changes sharply, and the miscalibration is substantial under each of them. Under SBM, HMC $95\,\%$ coverage on the three hidden prevalences ranges from $0.42$ (for $\log p_1$) to $0.76$ (for $\log p_3$), against a nominal $0.95$ and an interval-NBE $95\,\%$ PI coverage range of $0.945$ to $0.975$ on the same datasets. Coverage on the degree mean $\mu_d$ is closer to nominal under HMC ($0.940$) because the SBM specification used here distributes the known populations uniformly across blocks, so degree estimation is anchored by the conditional-independence assumption holding for the known-population responses. Under latent-space clustering, HMC $95\,\%$ coverage ranges from $0.70$ for $\sigma_d$ to $0.92$ for $p_3$, against interval-NBE coverage of $0.890$ to $0.985$. Under recall-subset thinning, HMC $95\,\%$ coverage is essentially zero on $\mu_d$ ($0.010$) and ranges from $0.61$ to $0.93$ on the remaining parameters, while the interval-NBE remains within $0.07$ of nominal across all five parameters.

\begin{table}[!t]
  \centering
  \caption{Empirical coverage of central credible intervals on 200 prior-predictive datasets per generative model.}
  \label{tab:mcmc_nbe_coverage}
    \resizebox{\textwidth}{!}{%
  \begin{tabular}{llcccc}
    \toprule
    & & \multicolumn{3}{c}{MCMC (Binomial-independent)} & {Interval-NBE} \\
    \cmidrule(lr){3-5}\cmidrule(lr){6-6}
    Generative model & Parameter & 50\,\% & 80\,\% & 95\,\% & 95\,\% \\
    \midrule
    Stochastic block model & $\mu_d$ & 0.450 & 0.815 & 0.940 & 0.975 \\
     & $\sigma_d$ & 0.345 & 0.660 & 0.865 & 0.885 \\
     & $p_1$ & 0.165 & 0.305 & 0.420 & 0.965 \\
     & $p_2$ & 0.250 & 0.445 & 0.585 & 0.975 \\
     & $p_3$ & 0.285 & 0.570 & 0.760 & 0.945 \\
    \midrule
    Latent-space & $\mu_d$ & 0.360 & 0.590 & 0.790 & 0.985 \\
     & $\sigma_d$ & 0.340 & 0.560 & 0.700 & 0.955 \\
     & $p_1$ & 0.315 & 0.550 & 0.725 & 0.890 \\
     & $p_2$ & 0.415 & 0.650 & 0.810 & 0.985 \\
     & $p_3$ & 0.470 & 0.740 & 0.915 & 0.975 \\
    \midrule
    Recall-subset & $\mu_d$ & 0.005 & 0.010 & 0.010 & 0.935 \\
     & $\sigma_d$ & 0.285 & 0.490 & 0.605 & 0.955 \\
     & $p_1$ & 0.420 & 0.650 & 0.815 & 0.975 \\
     & $p_2$ & 0.435 & 0.750 & 0.850 & 0.975 \\
     & $p_3$ & 0.435 & 0.720 & 0.925 & 0.960 \\
    \bottomrule
  \end{tabular}
}
\end{table}

% Discard rates (for caption or inline note):
%   standard: 0 / 200 discarded (0.000\%).
%   sbm: 0 / 200 discarded (0.000\%).
%   latent: 0 / 200 discarded (0.000\%).
%   recall: 0 / 200 discarded (0.000\%).

Taken together, the benchmark provides a quantitative measure of the cost of the conditional-independence assumption. It compares a correctly-specified-for-its-own-assumptions HMC, not a strawman implementation, against the framework on data that violate those assumptions, and on which no tractable alternative likelihood is available. The finding is that the likelihood-based approach is materially miscalibrated under all three forms of dependence, that the framework remains near-nominal in each, and that the miscalibration pattern is diagnostic of which form of dependence the data actually exhibit. A per-DGP interpretation of where in parameter space the conditional-independence likelihood absorbs the misspecification is reported in the Supplementary Material.

\subsection{Detailed per-DGP performance and prior-sensitivity}
\label{sec:sim_details_supplement}

Per-DGP performance is reported in full in the Supplementary Material;
we summarise the headline findings here. On $1{,}000$ prior-predictive
draws per DGP, median recovery is strong: Pearson correlation against
the truth exceeds $0.97$ on all five parameters under SBM and
latent-space, and exceeds $0.97$ on the three log-prevalences under
recall-subset. Empirical $95\%$ credible-interval coverage lies
within $0.04$ of nominal for every parameter under SBM and latent-space.
Under recall-subset the framework correctly detects the structural
confound between $\mu_d$ and the recall fraction $q_i$ and widens the
$\mu_d$ credible interval roughly fivefold relative to SBM and
latent-space, with coverage on $\mu_d$ remaining near nominal at $0.949$;
coverage on the log-prevalences remains within $0.03$ of nominal.

Sensitivity to prior misspecification on the hidden prevalences is
assessed in the Supplementary Material by re-evaluating each trained
NBE under two perturbed priors that preserve the means but inflate
variance and tail mass. Coverage degrades smoothly rather than
catastrophically (13 of 15 parameter rows above $0.90$ under the wider
perturbation; 10 of 15 under the broader one), and the degradation
concentrates in the lowest-prevalence hidden population $\log p_3$.
The diagnostic implication is that a practitioner running the framework
on a survey expected to lie in the tail of the training prior should
retrain on a prior that brackets the target regime.

\section{Application: Rwanda Estimating the Size of Populations through a Household Survey}
\label{sec:rwanda}

\subsection{Survey and data description}

We apply the framework to data from the Rwanda Estimating the Size of Populations through a Household Survey (ESPHS), conducted in 2011 by the Rwanda Biomedical Center in collaboration with the National Institute of Statistics of Rwanda. The ESPHS was designed to assess the feasibility of the network scale-up and proxy respondent methods for estimating the sizes of key populations at higher risk of HIV in Rwanda. The full ESPHS technical report \citep{rwanda2012esphs} documents the population definition, sample design, fielding dates, response rates with calculation details, and exact question wording. We access the public-use dataset through the \textsf{networkreporting} R package \citep{feehan2016networkreporting}, which also provides the design and analytic weights used here. The survey collected counts for 22 known populations whose sizes are verifiable from administrative records, including demographic groups, occupational categories, and common Rwandan first names. The hidden populations of interest are female sex workers (FSW), men who have sex with men (MSM), people who inject drugs (IDU), and clients of female sex workers.

\subsection{Prior calibration for the Rwanda setting}

We parametrise the training priors for the Rwanda application as follows. Hidden prevalences are parametrised in $\log_{10}$ space with $\log_{10} p_k \sim \mathrm{Uniform}(-5.0, -1.5)$, which covers the full ESPHS range. Degree priors are widened to $\mu_d \sim \mathrm{Uniform}(3, 5)$ and $\sigma_d \sim \mathrm{Uniform}(0.5, 2.5)$ to cover the Killworth anchor comfortably. The remaining nuisance priors are as in Section~\ref{sec:models}. A separate interval-NBE is trained for each of the three intractable models under these priors, using an encoder that had a 128-unit inner network, depth 3, and encoding dimension 64 to accommodate the broader prior range. We then evaluate each trained estimator on the ESPHS counts to obtain a posterior median and a $95\%$ credible interval for each parameter under each model.

\subsection{Comparative posterior inferences}

\begin{figure}[t]
\centering
\includegraphics[width=\textwidth]{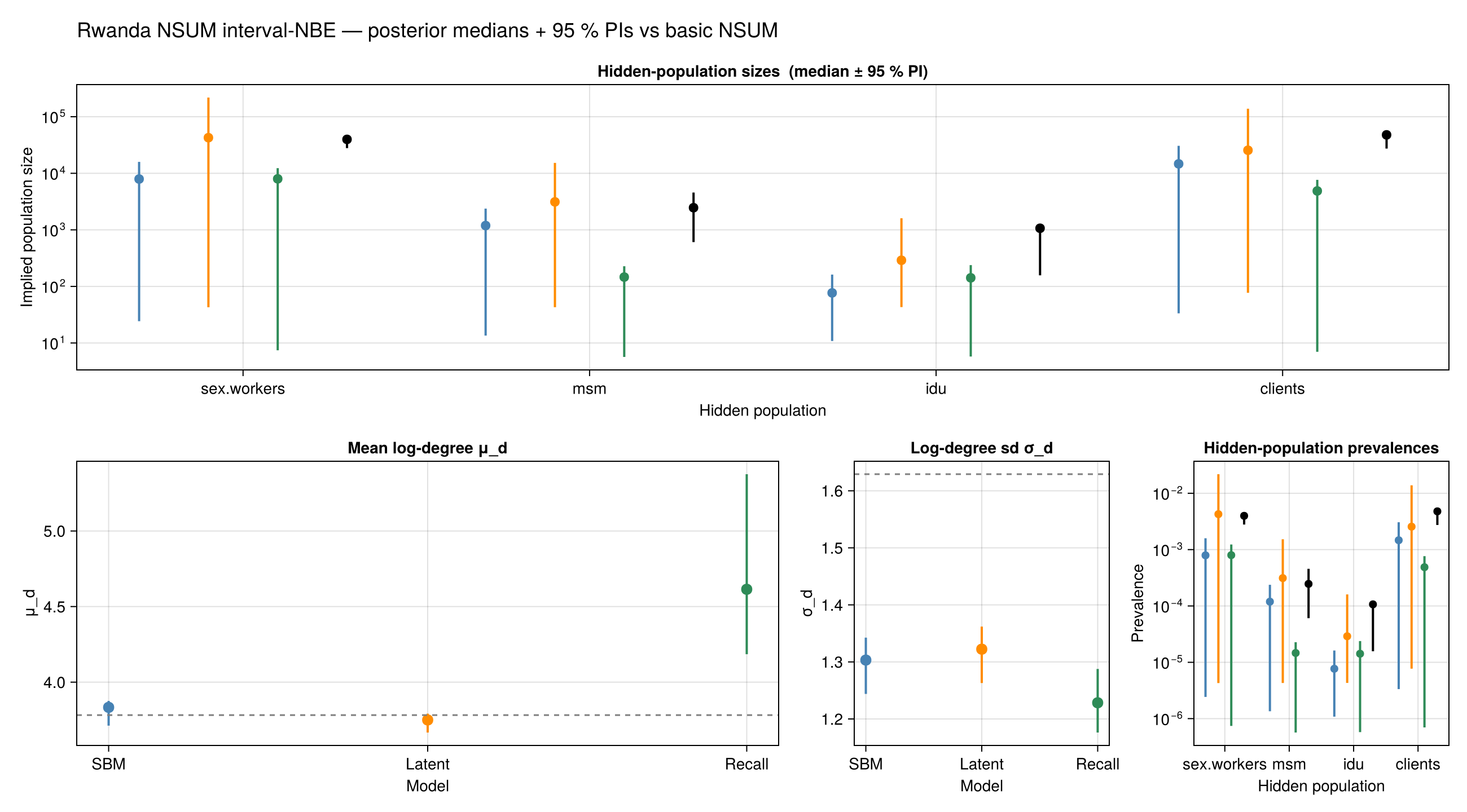}
\caption{Posterior medians and $95\%$ credible intervals for the four hidden prevalences [SBM (blue), Latent Space (orange), Recall (green)] under each of the three intractable models, with the basic NSUM Killworth estimator (black) and its rescaled-bootstrap $95\%$ confidence interval shown for comparison.}
\label{fig:rwanda_comparison}
\end{figure}

Table~\ref{tab:rwanda_results} reports the implied hidden-population sizes under each of the three intractable models, alongside the basic NSUM estimator as a likelihood-based benchmark. Figure~\ref{fig:rwanda_comparison} displays the same information graphically, with each model's posterior median and $95\%$ credible interval shown side by side per population. The headline finding is one of structural sensitivity. The three intractable models produce point estimates that disagree by factors of two to ten across the four hidden populations, which is the substantive analogue of the calibration sensitivity established in the simulation study. Where the survey signal is strongest, on sex workers and clients, the latent-space model produces estimates broadly consistent with basic NSUM, while SBM tracks at roughly half the basic NSUM scale and the recall-subset model an order of magnitude below. On MSM the latent-space and basic NSUM point estimates again agree closely, with SBM at roughly half and recall-subset an order of magnitude below. On the smallest hidden population, IDU, all three intractable models produce estimates between roughly $80$ and $300$, against the basic NSUM estimate of $1{,}068$.
 
\begin{table}[!t]
  \centering
  \caption{Rwanda NSUM hidden-population size estimates.}
  \label{tab:rwanda_results}
  \small
  \setlength{\tabcolsep}{4pt}
  \renewcommand{\arraystretch}{1.15}
  \begin{tabular}{l r r r r}
    \toprule
    Estimator & sex.workers & msm & idu & clients \\
    \midrule
    SBM           & 7,920             & 1,192           & 77            & 14,680            \\
                  & {[}24,~15,908{]}  & {[}14,~2,374{]} & {[}11,~162{]} & {[}33,~30,602{]} \\
    \addlinespace[2pt]
    Latent-space  & 42,613            & 3,123           & 290           & 25,563            \\
                  & {[}43,~218,316{]} & {[}43,~15,272{]} & {[}43,~1,603{]} & {[}77,~138,505{]} \\
    \addlinespace[2pt]
    Recall-subset & 7,981             & 146             & 142           & 4,875             \\
                  & {[}7,~12,338{]}   & {[}6,~227{]}    & {[}6,~238{]}  & {[}7,~7,657{]}    \\
    \addlinespace[2pt]
    Basic NSUM    & 39,791            & 2,464           & 1,068         & 47,798            \\
                  & {[}27,870,~41,513{]} & {[}607,~4,563{]} & {[}157,~1,234{]} & {[}27,361,~43,056{]} \\
    \bottomrule
  \end{tabular}
\end{table}

The substantive implication is that the choice of dependence model matters as much as the choice of survey or estimator family. A practitioner reporting a single point estimate from any one of these models would obscure structural uncertainty that is genuinely present in the data and that no amount of additional sampling can resolve. The conditional-independence likelihood embedded in basic NSUM is one specific modelling choice; the three intractable models studied here are three alternatives that are coherent with what is known about acquaintance networks and survey response. We do not claim that any one of the three is the right model for ESPHS. The disagreement among them, together with the simulation evidence in Section~\ref{sec:simulation} that all three are well-calibrated for prior-predictive draws within their training distributions, is itself the principal finding of the application.
 
Three caveats temper the comparison. First, the basic NSUM bootstrap CIs for sex workers and clients have upper bounds slightly below the point estimates ($41{,}513 < 39{,}791$ and $43{,}056 < 47{,}798$). This is a known pathology of the percentile bootstrap on heavily skewed estimators and is a property of the benchmark, not of the framework.%\footnote{The percentile bootstrap is well known to misbehave when the bootstrap distribution of the estimator is heavily skewed; see e.g.\ \citet{hall1988bootstrap}.} 
Second, the framework's lower-bound credible intervals are pinned at the prior boundary $\log_{10} p = -5$ in several places (translating to population sizes of $5$ to $30$). This is a direct consequence of the deliberately wide log-uniform training prior we adopted to make the framework portable to settings where prevalence is genuinely unknown to within an order of magnitude; the network is honestly reporting that the data do not exclude prevalences as low as the prior allows. Tightening the prior to a range informed by domain knowledge would tighten the lower-bound PIs proportionally, at the cost of making the trained estimator usable only for surveys consistent with that tighter prior. Third, basic NSUM is not a ground truth. It is a benchmark that uses the same conditional-independence likelihood the framework was designed to relax.

\subsection{Lessons for practice}

The Rwanda analysis demonstrates the operational value of the framework. With three intractable generative models that no closed-form likelihood approach can fit, we obtain posterior median estimates and credible intervals for hidden-population sizes in milliseconds per survey once the estimators are trained. The flexibility this affords over model design is substantial. A practitioner is no longer restricted to whichever extensions of the conditional-independence likelihood admit a tractable posterior; they can specify any generative model for which they can write a simulator and obtain calibrated inference within the framework's prior support. The same framework handles all three models in the same way, with no change to the inferential machinery.

The framework does not, however, tell the practitioner which model to choose. The disagreement among the three Rwanda estimates is not resolved by the framework itself; it is exposed by it. Choosing among the three competing models requires substantive input that the survey alone cannot supply, including domain knowledge about the network's dependence structure, evidence from validation studies in similar settings, or external information about the hidden populations of interest. The framework's contribution is to make the space of plausible models computationally accessible; the question of which model best describes the data-generating process for a particular survey remains a substantive one, and our application illustrates that it can matter by factors of two to ten.

\section{Discussion}
\label{sec:discussion}

We have developed an amortised neural estimation framework for Bayesian inference on aggregated relational data that separates inference from the need for a tractable likelihood, and we have demonstrated it on three structurally distinct intractable extensions of NSUM-style ARD inference. The same pipeline, an exchangeability-respecting DeepSets encoder feeding a multi-quantile head with a cumulative-gap construction, delivers near-nominal $95\%$ credible-interval coverage on all three cases, while MCMC on the conditional-independence likelihood applied to the same data is materially miscalibrated.

Several limitations warrant discussion. First, each of the three case studies captures one class of dependence; real social networks plausibly exhibit several simultaneously, alongside degree-degree correlations, triadic closure, transmission bias, and response heaping. These could be incorporated into a richer simulator, but each additional nuisance dimension costs training examples to marginalise. Second, the near-nominal coverage of Section~\ref{sec:simulation} is a property of the framework applied to three specific specifications and priors, not a generic property of the framework itself. Any new application requires its own modelling justification and its own coverage validation. We view this as a feature rather than a bug: the framework provides a principled pipeline for validating new assumptions, not a black-box estimator that obviates them. Third, where auxiliary information about community membership, geographic location, or interview conditions is available, it could be used as a covariate to reduce the effective dimensionality of the nuisance integration; we have assumed fully latent structure throughout. Fourth, the amortised nature of neural estimation means separate estimators are needed for substantially different survey designs. Fifth, as the Rwanda application illustrates, the method is sensitive to prior-data mismatch, and prior calibration from domain knowledge or pilot data is essential in practice.

Several alternatives to simulation-based neural inference exist for intractable likelihoods. Pseudo-marginal MCMC \citep{andrieu2009pseudo, andrieu2010particle} retains exact posterior targeting at the cost of nested Monte Carlo over block labels, latent positions, or per-respondent recall fractions, which is expensive and hard to tune in our settings. Data augmentation \citep{tanner1987calculation} can in principle render the latents conditionally tractable in all three models, but the weak per-respondent information about them makes mixing unreliable. Variational inference \citep{blei2017variational} requires a bespoke family per model and an approximation that is hard to audit without a tractable likelihood. Approximate Bayesian computation \citep{beaumont2002approximate, marin2012abc} avoids the likelihood but is not amortised; composite-likelihood approaches \citep{varin2011overview} re-introduce the independence shortcut the framework is designed to avoid. The combination of amortisation and likelihood-free training distinguishes the neural framework for ARD generative models with genuinely intractable likelihoods.

A natural further direction is full posterior inference. The interval-NBE deliberately reduces the inferential target to a posterior median and a $95\%$ credible interval per marginal parameter, which suffices for most reporting in survey statistics but does not provide joint posterior samples or higher-order summaries. We considered training a neural posterior estimator via a conditional normalising flow on the same DeepSets encoder, but found it difficult to calibrate in a principled manner on these simulators. Rank-histogram diagnostics revealed residual miscalibration that did not respond reliably to ensembling, training-set scaling, or flow-architecture changes. A robust workflow for calibrating amortised density estimators in the presence of high-dimensional latent nuisance is the natural follow-up to the present work. A complementary route is to apply post-hoc conformalised adjustment to the interval-NBE outputs \citep{angelopoulos2021gentle}, delivering finite-sample marginal coverage at the cost of losing the heteroscedasticity captured by the quantile head.

Within the framework, richer generative models are immediate: barrier effects \citep{maltiel2015estimating}, transmission bias \citep{mccormick2010many}, response heaping, and combinations of the dependence mechanisms studied here. The most pressing open problem is model comparison. Training several intractable models against the same survey, as in the Rwanda application, gives a structural sensitivity analysis but does not identify which model best describes the data. The standard tools, Bayes factors, posterior predictive checks, and information criteria, all require either a tractable likelihood or posterior samples, neither of which the framework produces; cross-validated count prediction is the natural fallback but is not aligned with the inferential target of population-level prevalences. The Rwanda application makes the urgency of this concrete: with no principled way to adjudicate between the three intractable models from the survey data alone, the practitioner is left with a structural sensitivity analysis but no model selection. Developing model-comparison tools within the simulation-based framework, perhaps via amortised model-evidence estimation or simulator-based goodness-of-fit checks, is the most important methodological extension this work demands.

Although the case studies all target NSUM-style hidden-population estimation, the framework's machinery is not specific to that inferential target; it takes a permutation-invariant aggregation of count vectors and produces marginal-quantile estimates while marginalising over high-dimensional latent nuisance, which is the structure of ARD inference generally. Three classes of non-NSUM ARD application are immediate: personal-network-size estimation under generative models that go beyond log-normal degree distributions \citep{mccormick2010many}; latent network structure inference, in which the targets are properties of the underlying acquaintance graph \citep{zheng2006many}; and joint estimation of barrier and transmission effects alongside hidden-population sizes \citep{maltiel2015estimating}. In each case the conditional-independence likelihood familiar from NSUM is also the working assumption, and in each case a generative simulator is straightforward to write while the marginal likelihood is intractable.

Simulation-based neural estimation provides a principled and flexible approach to Bayesian inference on aggregated relational data. We believe this flexibility will be essential as ARD methods are applied to increasingly heterogeneous populations and to survey designs that expose more of the underlying network structure.

\section*{Data Availability Statement}

This study uses simulated data and the publicly available Rwanda ESPHS 2011 survey accessed through the \textsf{networkreporting} R package \citep{feehan2016networkreporting}. All code for data simulation, model training, and analysis is available at \url{https://github.com/HiddenHarmsHub/neural-nsum}.

\section*{Acknowledgements}

This work was supported by a UKRI Future Leaders Fellowship [MR/X034992/1].

\bibliographystyle{apalike}
\bibliography{references}

\end{document}

\maketitle

This supplementary document gives detailed per-DGP performance of the
interval-NBE on the three intractable extensions of NSUM
(Sections~\ref{sec:supp_sbm}--\ref{sec:supp_recall}) and the
sensitivity analysis under prior misspecification
(Section~\ref{sec:supp_ood}). The interval-NBE introduced and trained
in the main paper is evaluated here on $N = 1{,}000$ held-out
prior-predictive datasets per data-generating process. All notation,
priors, training protocol, and figure conventions follow the main
paper.

\section{Stochastic block model NSUM}
\label{sec:supp_sbm}
\label{sec:sim_sbm}

Table~\ref{tab:sbm_crosspop} reports the Pearson correlation matrix of per-tie rates $y_{ik}/d_i$ across the three hidden populations for a representative SBM draw ($n = 2{,}000$, $\rho = 10$, moderate block-weight concentration). The off-diagonals are uniformly elevated relative to the Poisson-independent baseline (in which they would be close to zero), confirming that the simulator produces the cross-population dependence that is the signature of latent block structure.

\begin{table}[t]
\centering
\caption{Cross-population correlation of per-tie rates $y_{ik}/d_i$ under a representative SBM-NSUM draw with strong block structure ($\rho = 10$).}
\label{tab:sbm_crosspop}
\begin{tabular}{lc}
\toprule
Population pair & Correlation \\
\midrule
Pop.~1, Pop.~2 & $+0.34$ \\
Pop.~1, Pop.~3 & $+0.29$ \\
Pop.~2, Pop.~3 & $+0.41$ \\
\bottomrule
\end{tabular}
\end{table}

Table~\ref{tab:sbm_intnbe} reports performance of the interval-NBE on $N = 1{,}000$ held-out SBM-NSUM datasets. The recover the true parameters well, the degree mean $\mu_d$ has Pearson correlation $0.997$ against the truth. All three log-prevalence parameters have correlation about $0.98$, although the NBE does struggle with the lowest prevalence parameters for group 3, the smallest group. The degree heterogeneity $\sigma_d$ has correlation $0.973$, slightly weaker because variability in the effective prevalence across blocks is partly absorbed into the empirical spread of counts. Empirical $95\%$ coverage lies within $0.04$ of nominal for every parameter, ranging from $0.913$ for $\sigma_d$ to $0.991$ for $\mu_d$.

\begin{table}[t]
\centering
\caption{Interval-NBE performance under SBM-NSUM, on $N = 1{,}000$ held-out test datasets. Bias, MAE, and mean PI width are on the parameter's native scale (log scale for $p_k$).}
\label{tab:sbm_intnbe}
\begin{tabular}{lccccc}
\toprule
& \multicolumn{3}{c}{Median recovery} & \multicolumn{2}{c}{$95\%$ PI} \\
\cmidrule(lr){2-4}\cmidrule(lr){5-6}
Parameter & Bias & MAE & Pearson corr.\ & Coverage & Mean width \\
\midrule
$\mu_d$       &  $\phantom{-}0.008$ & $0.029$ & $0.997$ & $0.991$ & $0.20$ \\
$\sigma_d$    &  $\phantom{-}0.038$ & $0.048$ & $0.973$ & $0.913$ & $0.21$ \\
$\log p_1$    &  $\phantom{-}0.043$ & $0.102$ & $0.985$ & $0.972$ & $0.75$ \\
$\log p_2$    & $-0.009$            & $0.100$ & $0.982$ & $0.977$ & $0.67$ \\
$\log p_3$    &  $\phantom{-}0.040$ & $0.168$ & $0.983$ & $0.951$ & $1.77$ \\
\bottomrule
\end{tabular}
\end{table}

\begin{figure}[t]
\centering
\includegraphics[width=\textwidth]{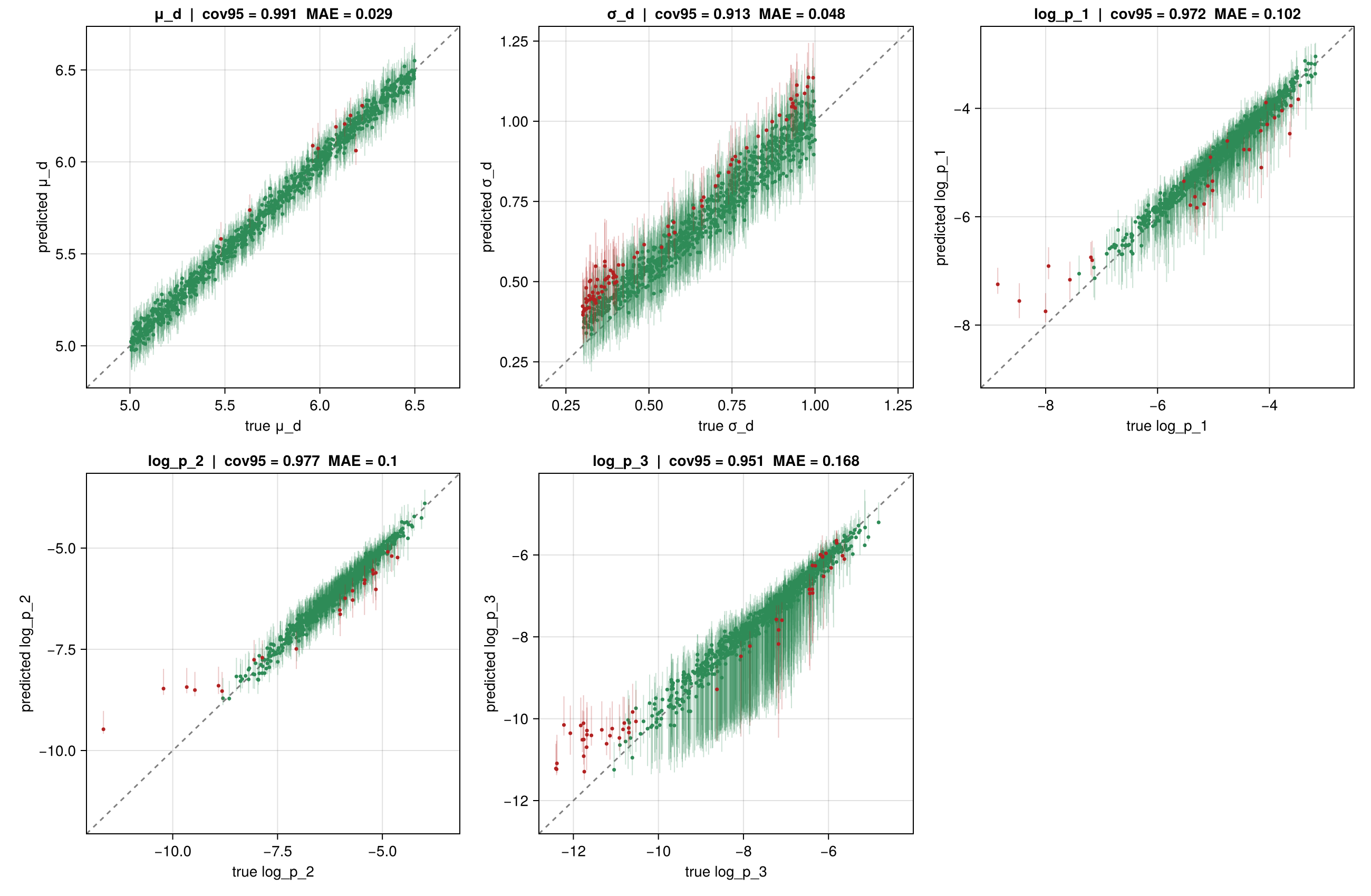}
\caption{Interval-NBE predictions on $N = 1{,}000$ held-out SBM-NSUM test datasets. Vertical bars are $95\%$ prediction intervals. Points are coloured green when the true value is contained in the interval and red otherwise.}
\label{fig:sbm_intnbe}
\end{figure}

Figure~\ref{fig:sbm_intnbe} shows scatter plots of Interval-NBE predictions against the ground truth across the held-out SBM-NSUM test sets. The reporting of median quality and interval calibration is informative on its own. The central estimate is, on average, very close to the true value (the median sits on the diagonal in Figure~\ref{fig:sbm_intnbe}), and the reported uncertainty is honest --- roughly $95\%$ of the true values fall inside the interval the estimator labels as a $95\%$ interval, so the estimator is neither overconfident (intervals that are too narrow) nor underconfident (intervals that are too wide). The largest mean width, $1.77$ on $\log p_3$, corresponds to the lowest-prevalence hidden population, where a typical respondent reports few or zero contacts and the data are correspondingly less informative. The estimator reflects this by widening the interval rather than over-claiming precision, which is the desired behaviour.

\section{Latent-space NSUM}
\label{sec:supp_latent}
\label{sec:sim_latent_space}

Table~\ref{tab:latent_crosspop} reports the cross-population correlation of per-tie rates alongside the corresponding latent-space distance between population centres, under sharp clustering ($\tau = 0.8$). The sign and magnitude of each off-diagonal track the geometry precisely. Close pairs show strong positive correlation, moderately distant pairs show weaker positive correlation, and pairs more than two bandwidths apart show negative correlation.

\begin{table}[t]
\centering
\caption{Cross-population correlation of per-tie rates $y_{ik}/d_i$ under sharp latent-space clustering, with corresponding centre distances $\|\mu_k - \mu_j\|$.}
\label{tab:latent_crosspop}
\begin{tabular}{lcc}
\toprule
Population pair & $\|\mu_k - \mu_j\|$ & Correlation \\
\midrule
Pop.~2, Pop.~3 & $0.60$ & $+0.66$ \\
Pop.~1, Pop.~3 & $1.07$ & $+0.28$ \\
Pop.~1, Pop.~2 & $1.64$ & $-0.11$ \\
\bottomrule
\end{tabular}
\end{table}

Table~\ref{tab:latent_intnbe} reports interval-NBE performance on $N = 1{,}000$ held-out latent-space NSUM datasets. Median recovery is strong across all five parameters: Pearson correlation exceeds $0.97$ for the log-prevalences and $0.98$ for the degree parameters. Empirical $95\%$ coverage lies within $0.04$ of nominal for every parameter, ranging from $0.918$ for $\log p_1$ to $0.980$ for $\mu_d$. Mean prediction-interval widths are tight on the degree parameters, comparable to SBM, and somewhat wider on the lowest-prevalence log-prevalence ($\log p_3$, mean width $2.11$).

\begin{table}[t]
\centering
\caption{Interval-NBE performance under latent-space NSUM, on $N = 1{,}000$ held-out test datasets. Bias, MAE, and mean PI width are on the parameter's native scale (log scale for $p_k$).}
\label{tab:latent_intnbe}
\begin{tabular}{lccccc}
\toprule
& \multicolumn{3}{c}{Median recovery} & \multicolumn{2}{c}{$95\%$ PI} \\
\cmidrule(lr){2-4}\cmidrule(lr){5-6}
Parameter & Bias & MAE & Pearson corr.\ & Coverage & Mean width \\
\midrule
$\mu_d$       &  $\phantom{-}0.017$ & $0.033$ & $0.996$ & $0.980$ & $0.22$ \\
$\sigma_d$    &  $\phantom{-}0.007$ & $0.032$ & $0.981$ & $0.978$ & $0.19$ \\
$\log p_1$    & $-0.064$            & $0.099$ & $0.972$ & $0.918$ & $0.75$ \\
$\log p_2$    & $-0.042$            & $0.096$ & $0.973$ & $0.975$ & $0.65$ \\
$\log p_3$    &  $\phantom{-}0.010$ & $0.154$ & $0.977$ & $0.950$ & $2.11$ \\
\bottomrule
\end{tabular}
\end{table}

\begin{figure}[t]
\centering
\includegraphics[width=\textwidth]{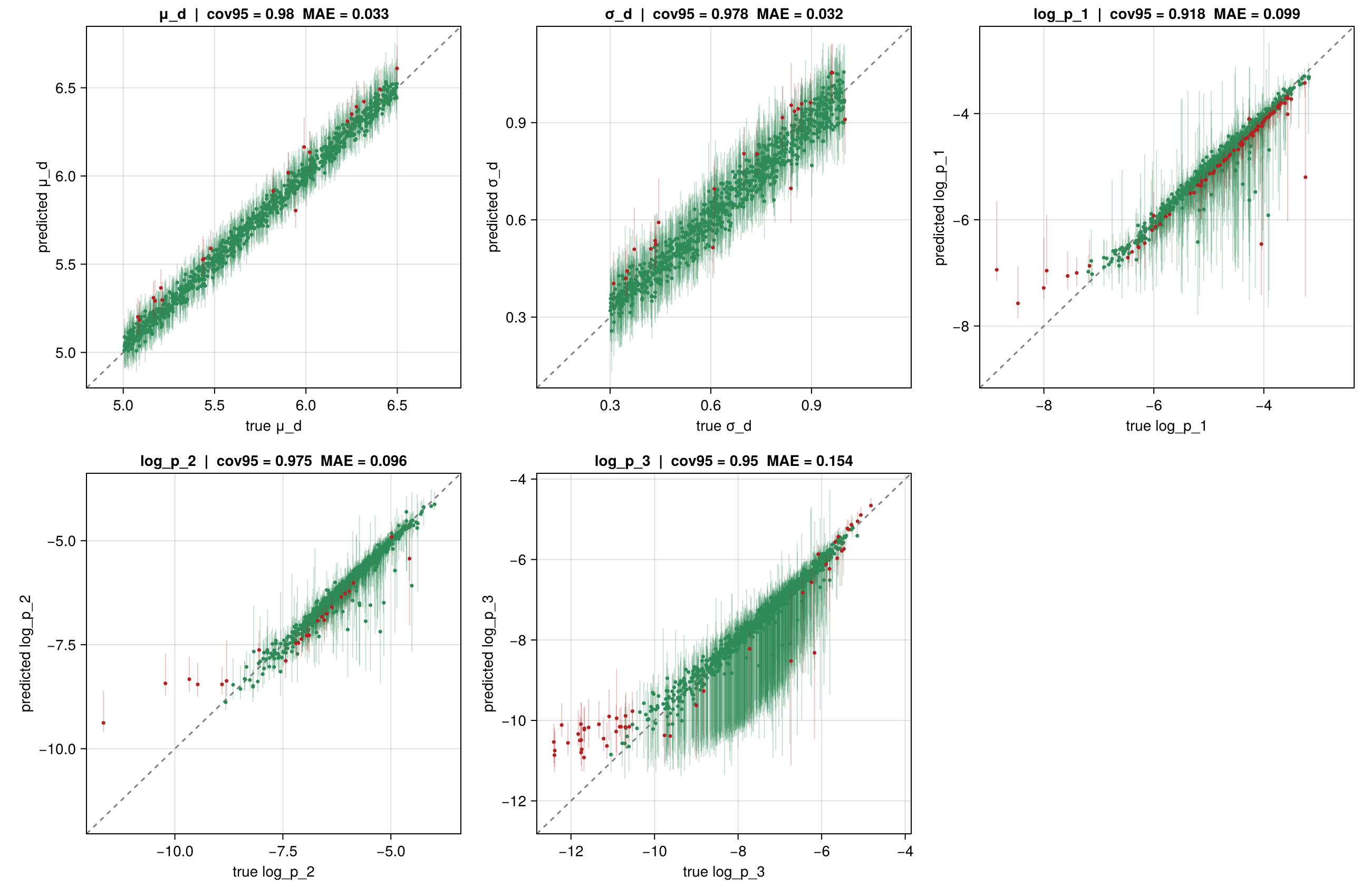}
\caption{Interval-NBE predictions on $N = 1{,}000$ held-out latent-space NSUM test datasets. Vertical bars are $95\%$ prediction intervals. Points are coloured green when the true value is contained in the interval and red otherwise.}
\label{fig:latent_intnbe}
\end{figure}

Figure~\ref{fig:latent_intnbe} reports the analogous diagnostic for the latent-space generator. The continuous high-dimensional latent structure with rotation and reflection symmetries, which is the feature that makes likelihood-based inference hardest for this model, passes through the framework without a visible penalty. The estimator targets $\theta$ directly, and the $O(n)$ latent coordinates are simply another source of nuisance variation the network absorbs at training time. Both parts of the diagnostic, median recovery and interval coverage, are met.

\section{Recall-subset NSUM}
\label{sec:supp_recall}
\label{sec:sim_recall}

Table~\ref{tab:recall_crosspop} reports the cross-population correlation under heterogeneous recall ($m_q = 0.5$, $\kappa_q = 3$) and near-uniform recall ($m_q = 0.8$, $\kappa_q = 200$). The off-diagonals are uniformly positive in the heterogeneous case (around $+0.46$ to $+0.48$) and collapse to near zero in the uniform case. A positive, uniform cross-population correlation structure is the signature of shared per-respondent thinning and is not expressible under the Poisson-independent likelihood. The variance-to-mean ratio is approximately 3 across all three populations under heterogeneous recall, against a Poisson baseline of 1, confirming the overdispersion.

\begin{table}[t]
\centering
\caption{Cross-population correlation of per-tie rates $y_{ik}/d_i$ under recall-subset NSUM. Heterogeneous: $(m_q, \kappa_q) = (0.5, 3)$. Near-uniform: $(0.8, 200)$.}
\label{tab:recall_crosspop}
\begin{tabular}{lcc}
\toprule
Population pair & Heterogeneous & Near-uniform \\
\midrule
Pop.~1, Pop.~2 & $+0.475$ & $-0.030$ \\
Pop.~1, Pop.~3 & $+0.478$ & $+0.016$ \\
Pop.~2, Pop.~3 & $+0.457$ & $+0.032$ \\
\bottomrule
\end{tabular}
\end{table}

Table~\ref{tab:recall_intnbe} reports interval-NBE performance on $N = 1{,}000$ held-out recall-subset NSUM datasets. The three log-prevalences are recovered with Pearson correlation above $0.97$ and median MAE between $0.05$ and $0.16$, comparable to or better than the SBM and latent-space cases. We find it difficult to recover the degree distribution parameters, $\mu_d$ has correlation $0.795$ and median MAE $0.218$, materially worse than under the other two simulators. This is a consequence of the data-generating process, as thinning and log-normal degree heterogeneity are partially confounded. A respondent with high $d_i$ and low $q_i$ is observationally similar to one with low $d_i$ and high $q_i$, and known-population anchors alone do not resolve the ambiguity.

Empirical $95\%$ coverage on $\mu_d$ remains near nominal at $0.949$, and the mean prediction-interval width on $\mu_d$ widens to $0.95$, more than four times wider than the corresponding values of $0.20$ and $0.22$ under SBM and latent-space respectively. This is to be expected, as we marginalise over the recall parameter $q$, and this parameter also reduces the amount of information in the data about the degree size parameters. The estimator quantifies the lost information by reporting a wider interval, rather than pretending to a precision the data do not support. Coverage on the log-prevalences remains within $0.03$ of nominal in every case.

\begin{table}[t]
\centering
\caption{Interval-NBE performance under recall-subset NSUM, on $N = 1{,}000$ held-out test datasets. Bias, MAE, and mean PI width are on the parameter's native scale (log scale for $p_k$).}
\label{tab:recall_intnbe}
\begin{tabular}{lccccc}
\toprule
& \multicolumn{3}{c}{Median recovery} & \multicolumn{2}{c}{$95\%$ PI} \\
\cmidrule(lr){2-4}\cmidrule(lr){5-6}
Parameter & Bias & MAE & Pearson corr.\ & Coverage & Mean width \\
\midrule
$\mu_d$       &  $\phantom{-}0.024$ & $0.218$ & $0.795$ & $0.949$ & $0.95$ \\
$\sigma_d$    &  $\phantom{-}0.003$ & $0.060$ & $0.914$ & $0.964$ & $0.36$ \\
$\log p_1$    & $-0.009$            & $0.051$ & $0.995$ & $0.981$ & $0.30$ \\
$\log p_2$    & $-0.001$            & $0.083$ & $0.987$ & $0.978$ & $0.50$ \\
$\log p_3$    &  $\phantom{-}0.016$ & $0.163$ & $0.977$ & $0.958$ & $2.14$ \\
\bottomrule
\end{tabular}
\end{table}

\begin{figure}[t]
\centering
\includegraphics[width=\textwidth]{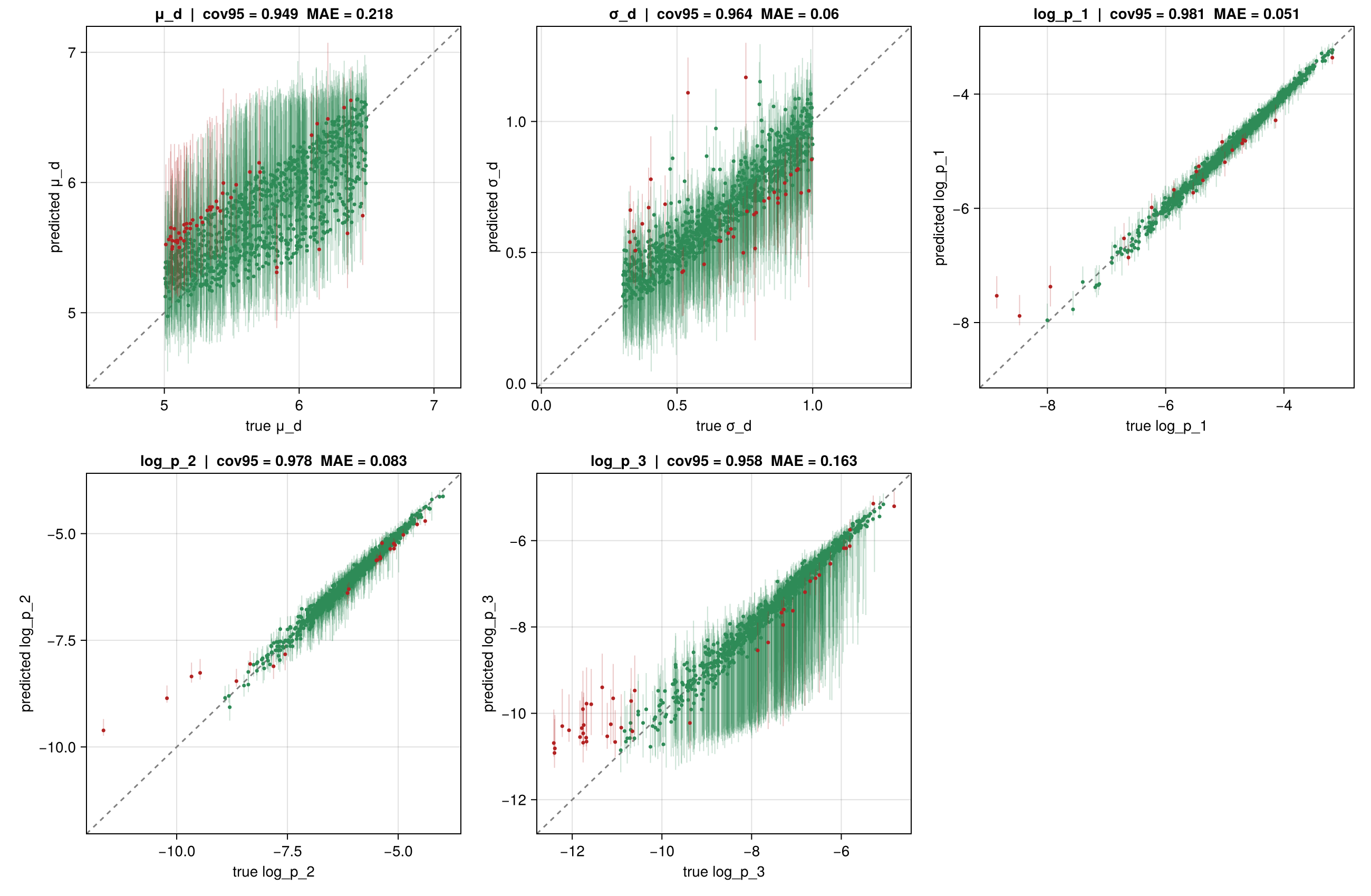}
\caption{Interval-NBE predictions on $N = 1{,}000$ held-out recall-subset NSUM test datasets. Vertical bars are $95\%$ prediction intervals. Points are coloured green when the true value is contained in the interval and red otherwise.}
\label{fig:recall_intnbe}
\end{figure}

Figure~\ref{fig:recall_intnbe} shows the corresponding scatter for the recall-subset generator. The Interval-NBE again recovers the targets with well-calibrated intervals, and the wider bars on the lowest-prevalence hidden populations reflect a smaller per-respondent count rather than estimator over-claiming.

\section{Sensitivity to prior misspecification}
\label{sec:supp_ood}
\label{sec:ood}

The framework is trained on a particular prior over hidden-population prevalences, and in any real application the practitioner will not know with certainty whether that prior brackets the regime from which the target survey is drawn. We assess sensitivity to prior misspecification by re-evaluating each trained interval-NBE on test sets drawn under two alternative priors on the hidden prevalences, treating the training prior as the reference and the alternatives as plausible misspecifications a practitioner might worry about. The training prior places independent $\mathrm{Beta}(2, 200)$, $\mathrm{Beta}(2, 600)$, and $\mathrm{Beta}(1, 1000)$ priors on $p_1$, $p_2$, and $p_3$ respectively. The \emph{wider} prior halves the Beta shape parameters so that the means are preserved but the variances roughly double; the \emph{broad} prior shifts the Beta scales to push more mass into the right tail. All other priors are held at their training values, and we draw $N = 1{,}000$ test datasets per misspecification per simulator.

Table~\ref{tab:intnbe_ood} reports the resulting median absolute error, $95\%$ prediction-interval coverage, and mean PI width for each parameter under each misspecification. Three patterns are worth surfacing. First, coverage degrades smoothly under misspecification rather than catastrophically: on the training prior, 13 of 15 parameter rows have coverage in $[0.94, 0.99]$; under the wider misspecification, 12 of 15 remain above $0.90$; under the broad misspecification, $10$ of 15 do. Second, the degradation concentrates in the lowest-prevalence hidden population, $\log p_3$, whose coverage drops from values close to $0.97$ on the training prior to between $0.85$ and $0.88$ under the misspecified priors across all three simulators. This is the regime in which the estimator has been trained on relatively little tail mass and the misspecified prior shifts substantial mass into precisely that tail. Third, the mean PI width is roughly stable across misspecifications while MAE grows substantially on the affected parameters. The intervals are not shrinking; the medians are drifting under prior misspecification, and the same intervals therefore fail to cover. This is the correct diagnostic interpretation, and it suggests a clear remedy: where the target survey's hidden-population prevalences are expected to lie in the tail of the training prior, the estimator should be retrained on a prior that brackets the target regime.

Degree parameters degrade more smoothly than hidden prevalences across all three simulators, which is consistent with degree estimation being anchored by known populations whose prevalences are held fixed under the misspecifications considered here. In consequence, any real-survey application of the framework should re-run the sensitivity check on candidate alternative priors that span the practitioner's uncertainty about the target regime, rather than relying on the in-distribution certificate alone.

\begin{table}[!t]
  \centering
  \caption{Sensitivity to prior misspecification on the hidden prevalences.}
  \label{tab:intnbe_ood}
  \resizebox{\textwidth}{!}{%
  \begin{tabular}{ll ccc ccc ccc}
    \toprule
    & & \multicolumn{3}{c}{Training prior} & \multicolumn{3}{c}{Wider misspecification} & \multicolumn{3}{c}{Broad misspecification} \\
    \cmidrule(lr){3-5}\cmidrule(lr){6-8}\cmidrule(lr){9-11}
    Generative model & Parameter & MAE & cov$_{95}$ & width & MAE & cov$_{95}$ & width & MAE & cov$_{95}$ & width \\
    \midrule
    Stochastic block model & $\mu_d$    & 0.029 & 0.990 & 0.201 & 0.031 & 0.974 & 0.196 & 0.039 & 0.929 & 0.194 \\
                           & $\sigma_d$ & 0.048 & 0.907 & 0.210 & 0.051 & 0.909 & 0.212 & 0.056 & 0.848 & 0.213 \\
                           & $\log p_1$ & 0.099 & 0.974 & 0.753 & 0.175 & 0.920 & 0.768 & 0.169 & 0.908 & 0.781 \\
                           & $\log p_2$ & 0.110 & 0.960 & 0.665 & 0.150 & 0.921 & 0.696 & 0.165 & 0.908 & 0.677 \\
                           & $\log p_3$ & 0.156 & 0.957 & 1.744 & 0.389 & 0.880 & 1.684 & 0.335 & 0.872 & 1.504 \\
    \midrule
    Latent-space          & $\mu_d$    & 0.033 & 0.973 & 0.217 & 0.036 & 0.949 & 0.217 & 0.042 & 0.904 & 0.218 \\
                          & $\sigma_d$ & 0.031 & 0.967 & 0.191 & 0.035 & 0.943 & 0.189 & 0.041 & 0.911 & 0.192 \\
                          & $\log p_1$ & 0.097 & 0.900 & 0.763 & 0.187 & 0.824 & 0.819 & 0.164 & 0.808 & 0.790 \\
                          & $\log p_2$ & 0.093 & 0.978 & 0.660 & 0.155 & 0.919 & 0.693 & 0.176 & 0.934 & 0.738 \\
                          & $\log p_3$ & 0.161 & 0.974 & 2.149 & 0.405 & 0.857 & 1.951 & 0.361 & 0.868 & 1.916 \\
    \midrule
    Recall-subset         & $\mu_d$    & 0.215 & 0.942 & 0.949 & 0.212 & 0.940 & 0.945 & 0.212 & 0.927 & 0.957 \\
                          & $\sigma_d$ & 0.058 & 0.955 & 0.359 & 0.064 & 0.950 & 0.392 & 0.068 & 0.928 & 0.366 \\
                          & $\log p_1$ & 0.050 & 0.983 & 0.296 & 0.094 & 0.941 & 0.337 & 0.092 & 0.919 & 0.319 \\
                          & $\log p_2$ & 0.074 & 0.977 & 0.483 & 0.146 & 0.906 & 0.507 & 0.141 & 0.901 & 0.469 \\
                          & $\log p_3$ & 0.148 & 0.961 & 2.100 & 0.377 & 0.880 & 1.974 & 0.287 & 0.884 & 1.910 \\
    \bottomrule
  \end{tabular}%
  }
\end{table}

\section{Failure modes of HMC under each DGP}
\label{sec:supp_hmc_failure}

The main paper reports the headline coverage of HMC on the Binomial-independent NSUM versus the interval-NBE across the well-specified positive control and the three misspecified DGPs. Here we interpret where in parameter space the conditional-independence likelihood absorbs the misspecification in each case, expanding on Table~1 in the main paper.

Under SBM, the prevalence parameters are most affected, with coverage as low as $0.42$, while the degree parameters retain near-nominal coverage. This is consistent with the conditional-independence likelihood having no degree of freedom available to absorb cross-population dependence on the hidden prevalences while the degree parameters are anchored by uniformly-distributed known populations. Under latent-space clustering, miscalibration is more even across parameters but degree parameters remain harder hit than prevalences, which is consistent with the kernel normalisation in equation~(8) of the main paper preserving the marginal $p_k$ in expectation while distorting per-respondent activations. Under recall thinning, the hardest-hit parameter is the degree mean $\mu_d$ ($95\,\%$ coverage of $0.010$), which is where the likelihood absorbs the misspecification. HMC commits confidently to a degree distribution consistent with observed count rates under unit recall, rather than acknowledging the recall-degree confound that the framework correctly flags by widening the prediction interval. Prevalences are the least affected of the three cases under recall thinning, consistent with the ratio-based NSUM estimator being largely robust to multiplicative recall factors that apply equally to known and hidden populations.